# Crosstalk calibration of multi-pixel photon counters using coherent states

**Dmitry A. Kalashnikov, Si-Hui Tan, and Leonid A. Krivitsky\***

*Data Storage Institute, Agency for Science Technology and Research, 5 Engineering Drive 1, 117608 Singapore*
*Leonid_Krivitskiy@dsi.a-star.edu.sg*

**Abstract:** We present a novel method of calibration of crosstalk probability for multi-pixel photon counters (MPPCs) based on the measurement of the normalized second-order intensity correlation function of coherent light. The method was tested for several MPPCs, and was shown to be advantageous over the traditional calibration method based on the measurements of the dark noise statistics. The method can be applied without the need of modification for different kinds of spatially resolved single photon detectors.

**References and links:**

1. E. Knill, R. Laflamme, and G. J. Milburn, "A scheme for efficient quantum computation with linear optics," Nature (London) **409**, 46-52 (2001).
2. P. Kok, W. J. Munro, K. Nemoto, T. C. Ralph, J. P. Dowling, and G. J. Milburn, "Linear optical quantum computing with photonic qubits," Rev. Mod. Phys. **79**, 135-175 (2007).
3. J. L. O'Brein, "Optical quantum computing," Science **318**, 1567-1570 (2007).
4. M. Zukowski, A. Zeilinger, M. A. Horne, and A. K. Ekert, "Event-ready detectors" Bell experiment via entanglement swapping," Phys. Rev. Lett. **71**, 4287-4290 (1993).
5. L. Pezze and A. Smerzi, "Mach-Zehnder interferometry at the Heisenberg limit with coherent and squeezed-vacuum light," Phys. Rev. Lett. **100**, 073601 (2008).
6. S. Cova, A. Longoni, and A. Andreoni, "Towards picosecond resolution with single-photon avalanche diodes," Rev. Sci. Inst. **52**, 408-412 (1981).
7. D. Achilles, C. Silberhorn, C. Sliwa, K. Banaszek, and I. A. Walmsley, "Fiber assisted detection with photon-number resolution," Opt. Lett. **28**, 2387-2389 (2003).
8. M. J. Fitch, B. C. Jacobs, T. B. Pittman, and J. D. Franson, "Photon number resolution using a time-multiplexed single-photon detector," Phys. Rev. A **68**, 043814 (2003).
9. M. Mičuda, O. Haderka, and M. Ježek, "High-efficiency photon-number-resolving multichannel detector," Phys Rev A **78**, 025804 (2008).
10. M. Bondani, A. Allevi, A. Agliati, and A. Andreoni, "Self-consistent characterization of light statistics," J. Mod. Opt. **56,** 226-231 (2009).
11. S. Takeuchi, J. Kim, and Y. Yamamoto, "Multiphoton detection using visible light photon counter," Appl. Phys. Lett. **74**, 902-904 (1999).
12. J. Kim, S. Takeuchi, Y. Yamamoto, and H. H. Hogue, "Development of a high quantum-efficiency single-photon counting system," Appl. Phys. Lett. **74**, 1063–1065 (1999).
13. E. Waks, K. Inoue, E. Diamanti, and Y. Yamamoto, "High-efficiency photon-number detection for quantum information processing," IEEE J. Sel. Top. Quant. **9**, 1502-1511 (2003).
14. B. Cabrera, R. M. Clarke, P. Colling, A. J. Miller, S. Nam, and R. W. Romani, "Detection of single infrared, optical, and ultraviolet photons using superconducting transition edge sensors," Appl. Phys. Rev. **73**, 735-737 (1998).
15. D. Rosenberg, A. E. Lita, A. J. Miller, and S. W. Nam, "Noise-free high-efficiency photon-number-resolving detectors," Phys. Rev. A **71**, 061803R (2005).
16. A. E. Lita, A. J. Miller, and S. W. Nam, "Counting near-infrared single- photons with 95% efficiency," Opt. Exp. **16**, 3032-3040 (2008).
17. Hamamatsu web-page http://jp.hamamatsu.com/products/sensor-ssd/4010/index_en.html
18. I. Afek, A. Natan, O. Ambar, and Y. Silberberg, "Quantum state measurements using multipixel photon detectors," Phys. Rev. A **79**, 043830 (2009).
19. M. Ramilli, A. Allevi, V. Chmill, M. Bondani, M. Caccia, and A. Andreoni, "Photon-number statistics with silicon photomultipliers," JOSA B **27**, 852-862 (2010).

20. D. A. Kalashnikov, S.-H. Tan, M. V. Chekhova, and L. A. Krivitsky, "Accessing photon bunching with photon number resolving multi-pixel detector," Opt. Exp. **19**, 9352-9363 (2011).
21. L. Dovrat, M. Bakstein, D. Istrati, A. Shaham, H. S. Eisenberg, "Measurements of the dependence of the photon-number distribution on the number of modes in parametric down-conversion," Opt. Exp. **20**, 2266-2276 (2012).
22. A. Vacheret, G. J. Barker, M. Dziewiecki, P. Guzowski, M. D. Haigh, B. Hartfiel, A. Izmaylov, W. Johnston, M. Khabibullin, A. Khotjantsev, Yu. Kudenko, R. Kurjata, T. Kutter, T. Lindner, P. Masliah, J. Marzec, O. Mineev, Yu. Musienko, S. Oser, F. Retiere, R. O. Salih, A. Shaikhiev, L. F. Thompson, M. A. Ward, R. J. Wilson, N. Yershov, K. Zaremba, and M. Ziembicki, "Characterization and simulation of the response of multi pixel photon counters to low light levels," ArXiv: 1101.1996v1 (2011)
23. M. Akiba, K. Tsujino, K. Sato, and M. Sasaki, "Multipixel silicon avalanche photodiode with ultralow dark count rate at liquid nitrogen temperature," Opt. Exp. **17**, 16885-16897 (2009).
24. A. Spinelli and A. L. Lacaita, "Physics and numerical simulation of single photon avalanche diodes," IEEE Trans. on Electron Devices **44**, 1931-1943 (1997).
25. I. Rech, A. Ingargiola, R. Spinelli, I. Labanca, S. Marangoni, M. Ghioni, and S. Cova, "A new approach to optical crosstalk modeling in single-photon avalanche diodes," IEEE Phot. Tech. Lett. **20**, 330-332 (2008).
26. R. D. Younger, K. A. McIntosh, J. W. Chludzinski, D. C. Oakley, L. J. Mahoney, J. E. Funk, J. P. Donelly, and S. Verghese, "Crosstalk analsis of integrated Geiger-mode avalanche photodiode focal plane array," Proc. of SPIE **7320**, 73200Q-12 (2009).
27. E. Sciacca, G. Condorelli, S. Aurite, S. Lombardo, M. Mazzillo, D. Sanfilippo, G. Fallica, and E. Rimini, "Crosstalk characterization in Geiger-mode avalanche photodiode arrays," IEEE Electron Devices Letters **29**, 218-220 (2008).
28. M. Yokoyama, A. Minamino, S. Gomi, K. Ieki, N. Nagai, T. Nakaya, K. Nitta, D. Orme, M. Otani, T. Murakami, T. Nakadaira, and M. Tanaka, "Performance of multi-pixel photon counters for the T2K near detectors," ArXiv:1007.2712v1 (2010).
29. P. Eraerds, M. Legre, A. Rochas, H. Zbinden, and N. Gisin, "SiPM for fast photon-counting and multiphoton detection," Opt. Exp. **15**, 14539-14549 (2007).
30. M. Avenhaus, K. Laiho, M. V. Chekhova, and C. Silberhorn, "Accessing higher order correlations in quantum optical states by time multiplexing," Phys. Rev. Lett. **104,** 063602 (2010).
31. A. L. Lacaita, F. Zappa, S. Bigliardi, and M. Manfedi, "On the bremsstrahlung origin of hot-carrier-induced photons in silicon devices," IEEE Trans. on Electron Devices **40**, 577-582 (1993).

**1. Introduction**

Accurate characterization of multiphoton states is one of the key tasks of modern quantum optics. It requires implementation of photodetectors capable to resolve numbers of photons in short optical pulses, referred to as photon number resolving detectors (PNRDs). Possible applications of PNRDs include but not limited to linear optical quantum computing [1-3], security analysis of quantum key distribution schemes [4], and studies of foundations of quantum mechanics [5].

Conventional single photon detectors, such as avalanche photodiodes (APDs) and photomultiplier tubes (PMTs), lack the photon number resolving capability because of the "dead-time" effect [6]. Thus their output corresponds either to "zero" or to "one or more" detected photons. This limitation can be overcome by splitting the multiphoton state between different spatial or temporal modes and detecting each mode individually [7-9]. However, with the increasing number of photons these schemes become bulky and challenging to operate. Note that limited photon number resolution is offered by a dedicated class of hybrid photodetectors [10]. An alternative technique is based on the implementation of cryogenic devices such as transition edge sensors (TESs) and visible light photon counters (VLPCs). They have excellent photon number resolution up to 7-9 photons, high quantum efficiency and low dark noise [11-16]. However, these devices are costly and require qualified operation.

Recently, an affordable solution for PNRDs became available with the implementation of multi-pixel photon counters (MPPC), where several hundred APDs, operating in a Geiger mode and referred to as pixels, are embedded into a single chip [17]. The chip is illuminated by a diffused light spot, so that the probability of more than one photon to hit the same pixel is negligible. Signals from all the pixels are summarized at the output and thus the photon number resolution is achieved. In some sense this approach is similar to the above-mentioned technique of spreading multiphoton states in different spatial modes, being, however, more compact and easy to operate. Currently, MPPCs find applications in quantum optics experiments [18-21], as well as in high-energy physics experiments as photosensors for scintillate detectors [22].

The major drawbacks of MPPC are high dark noise and optical crosstalk between pixels. The dark noise can be significantly reduced by cooling MPPC and time gating of the signal [23]. However the optical crosstalk cannot be suppressed in this way and thus requires more elaborated attention. Crosstalk occurs because secondary photons are re-emitted during the avalanche in the pixel and they trigger simultaneous additional photon counts [24]. Thus the crosstalk modifies the initial statistics of photons, and that is why its accurate calibration is needed.

## 2. Methods of crosstalk calibration

The direct calibration of the crosstalk requires either the access to the outputs of individual pixels [25-27] which is not possible with commercially available MPPCs, or complicated optical setups which provide illumination of individual pixels [22].

Another widely implemented approach is based on the measurement of the photocount distribution of dark noise [22, 28]. Assuming that dark counts obey a Poisson distribution in the absence of crosstalk, the corresponding mean number of dark counts per measurement is given by:

$$\langle N \rangle_{DC} = -\ln(N_{0,DC}/N_{DC}), \tag{1}$$

where $N_{0,DC}$ is the number of measured events with zero counts, and $N_{DC}$ is the total number of measurements (triggers). The probability of the crosstalk is assumed to be proportional to the difference between the number of single photon events expected from the Poisson distribution with the mean $\langle N \rangle_{DC}$ and the corresponding number of events $N_{1,DC}$, measured for the dark noise:

$$p_{DC} = 1 - N_{1,DC} \bigg/ \left( N_{DC} \langle N \rangle_{DC} e^{-\langle N \rangle_{DC}} \right). \tag{2}$$

This approach allows calibration of the crosstalk without much effort. However, in case of low dark counts the difference between the nominator and the denominator in Eq. (2) is small and hence the crosstalk calibration is highly sensitive to the errors in the measurement of both terms. Thus the method does not allow accurate calibration of low-noise detectors, which are of most practical interest.

An alternative method of crosstalk calibration relies on statistical modeling of the measured photon statistics of well characterized light sources and solving an inverse problem. To date, several analytical models describing MPPC crosstalk have been considered [18-21, 29]. However, in the analysis of the experimental data either the probability of crosstalk has been analyzed as one of several fitting parameters [19] or the assumption has been made that for a given pixel only one crosstalk event could occur [18, 20, 29]. The latter is in fact valid, particularly, for the case when the number of photons, dark counts, and the probability of crosstalk are low. However for the cases when the probability of crosstalk is high, several crosstalk photons are likely to be generated in a single detection event and a nonlinear crosstalk model should be considered.

Recently, an algorithm of obtaining the normalized second order intensity correlation function ($g^{(2)}$) from the MPPC data was introduced [20]. Based on a simple linear model of crosstalk in MPPC it was shown, that due to the crosstalk, the measured $g^{(2)}$ exhibits excess two-photon correlations independently on the actual light statistics. In this work by implementing a more general non-linear crosstalk model, which is applicable to a broader class of MPPC devices, we develop a method of the crosstalk calibration based on $g^{(2)}$ measurements of coherent light and test it for several commercially available MPPCs.

### 3. Non-linear crosstalk model

According to the definition, the second-order normalized intensity correlation function at zero time delay is given by

$$g^{(2)} \equiv \langle a^{+2} a^2 \rangle / \langle a^+ a \rangle^2, \quad (3)$$

where $a^+$, $a$ are photon creation and annihilation operators, respectively. Measurement of $g^{(2)}$ is highly relevant to characterization of light statistics since it is insensitive to optical losses and finite efficiencies of the detectors, which may be not always precisely known in experiments [30]. Conventionally $g^{(2)}$ is measured in a Hanbury-Brown and Twiss (HBT) setup consisting of two single photon detectors, which outputs are addressed to a coincidence circuit. Let $N_{Coinc}$ be the measured number of coincidences of photocounts of two detectors, and $N_{D1,D2}$ is the measured number of photocounts of individual photodetectors, then $g^{(2)}$ is given by:

$$g^{(2)} = N_{Coinc} / (N_{D1} N_{D2}). \quad (4)$$

For the case of MPPC, one considers each pair of pixels as a single HBT setup [20]. The number of such setups can be estimated as $m(m-1)/2 \approx m^2/2$, where $m$ is the total number of pixels. Let $N_{k,CT}$ be the number of detected $k$-photon events by MPPC ($k=1,2,3...$), including crosstalk events. Then, the total number of pairwise coincidences is given by $N_{Coinc,CT} = \sum_{k=2}^{\infty} C_2^k N_{k,CT}$, where $C_2^k$ is the number of 2-combinations in $k$. The total amount of photocounts is given by $N_{Total,CT} = \sum_{k=1}^{\infty} k N_{k,CT}$. In analogy with Eq. (4), $g^{(2)}$ is calculated as a ratio of the number of pairwise coincidences per single HBT setup to the squared total number of detected photons per pixel:

$$g^{(2)} = 2 \sum_{k=2}^{\infty} C_2^k N_{k,CT} / \left( \sum_{k=1}^{\infty} k N_{k,CT} \right)^2. \quad (5)$$

Let us now introduce the nonlinear crosstalk model. One expresses the number of k-photon events with crosstalk $N_{k,CT}$ through the corresponding number of events in the absence of crosstalk $N_k$ as follows:

$$k=1 \quad N_{1,CT} = N_1 - p N_1 - p^2 N_1, \quad (6a)$$

$$k=2 \quad N_{2,CT} = N_2 - 2p N_2 + p N_1 - 2p^2 N_2, \quad (6b)$$

$$k>2 \quad N_{k,CT} = N_k - kp N_k + (k-1) p N_{k-1} + (k-2) p^2 N_{k-2} - k p^2 N_k, \quad (6c)$$

where $p$ is the probability of crosstalk for one pixel. In Eq. (6(c)) the second term is the number of $k$-photon events being converted by crosstalk into $(k+1)$-photon events, and the third term is the number of $(k-1)$-photon events that gained an extra photon due to crosstalk to become $k$-photon events; the fourth term represents the number of $(k-2)$-photon events converted into $k$-photon events due to double crosstalk, and the fifth term is the number of $k$-photon events which contributed to $(k+2)$-photon events, also due to double crosstalk. The model is restricted to the second order crosstalk under the assumption of low mean photon number of the detected light, and the following condition for the crosstalk probability to be fulfilled $p + 2p^2 >> 3p^3$. However the model can be extended to include further nonlinear terms in the similar manner.

Assuming the crosstalk model mentioned above, the nominator and the denominator in Eq. (5) are given by:

$$N_{Coinc,CT} = (1 + 2p + 4p^2) \sum_{k=2}^{\infty} C_2^k N_k + p(1 + 3p) \sum_{k=1}^{\infty} k N_k, \quad (7a)$$

$$N_{Total,CT} = (1 + p + 2p^2) \sum_{k=1}^{\infty} k N_k. \quad (7b)$$

From Eqs. (5, 7(a), 7(b)) $g^{(2)}$ takes the following form

$$g^{(2)}(N_{Total,CT}) = \frac{(1 + 2p + 4p^2)}{(1 + p + 2p^2)^2} g_0^{(2)} + \frac{2p(1 + 3p)}{(1 + p + 2p^2)} \frac{1}{N_{Total,CT}}, \quad (8)$$

where $g_0^{(2)} \equiv 2\sum_{k=2}^{\infty} C_2^k N_k / \left(\sum_{k=1}^{\infty} kN_k\right)^2$, is the initial second order correlation of light without crosstalk. From Eq. (8) it follows that for light with the known $g_0^{(2)}$, for instance for coherent light, for which $g_0^{(2)}=1$, the measured $g^{(2)}$ only depends on the crosstalk probability and the total number of photocounts. Thus, if the dependence of $g^{(2)}$ on the total number of photocounts $N_{Total,CT}$ is measured, the probability of crosstalk can be extracted as the only fitting parameter. Note that since $g_0^{(2)}$ is insensitive to optical losses, the described method of crosstalk calibration does not require *a-priori* knowledge of the quantum efficiency of the detector under test.

## 4. Experiment

The approach was tested experimentally in the setup shown in Fig.1. The beam of a frequency doubled Nd:YAG pulsed laser at 532 nm (Crystalaser, repetition rate 20 kHz, pulse width 30ns) was attenuated by an neutral density filter (NDF) and fed into a single mode (SMF) fiber (SM 460-HP). At the output of the fiber the beam was collimated by a lens with $f=6.24$ mm to produce a Gaussian spot of 1.2 mm at the level of $1/e^2$ from the maximum. The intensity of the beam was varied by two polarization beam splitters (PBS) and a half-wave plate (HWP) placed between them. The beam passed through an interference filter with a center wavelength 532 nm, FWHM 2 nm (not shown) and impinged on the chip of MPPC. Three different MPPC's (Hamamatsu, C10507-11-XXXU series) with 100, 400, 1600 pixels per chip with the corresponding pixel sizes of 100x100 μm², 50x50 μm², and 25x25 μm², were tested. The size of the MPPC chip was 1.5x1.5 mm² for all the detectors under test, and the quantum efficiencies at 25ºC according to the datasheet were equal to 58%, 38%, 20% for MPPCs with pixel sizes of 100x100 μm², 50x50 μm², and 25x25 μm², respectively. Each MPPC was sealed into custom-made housing with AR-coated optical window. The temperature of MPPC could be varied thermoelectrically from the room temperature down to -8ºC and monitored by a thermocouple. The signals from MPPC were digitized by an AD card (NI PCI-5154) within 50 ns time window, and subsequently processed by Labview software. For each value of the intensity of the impinging light distribution of the amplitudes of the MPPC output $N_{k,CT}$ was recorded for $2*10^6$ triggers. The dark noise was measured by blocking the laser beam, and then subtracted from the signal. The total amount of photocounts $N_{Total,CT}$ was calculated according to Eq. (7(b)), and $g^{(2)}$ was calculated according to Eq. (5). The experimental dependence of $g^{(2)}(N_{Total,CT})$ was fitted with Eq. (8) by a Levenberg-Marquardt algorithm weighted for instrumental uncertainties (Origin lab), with *p* being a single fitting parameter.

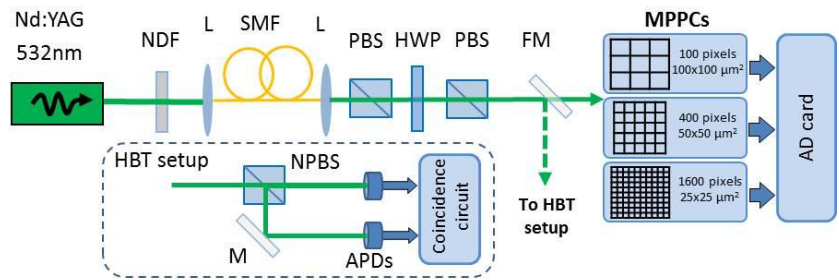

Fig.1 Experimental setup. Nd:YAG laser at 532 nm was used as a source of coherent light. The beam was attenuated by a neutral density filter (ND), and coupled into a single mode fiber (SMF). The intensity of the beam was controlled by two polarizing beamsplitters (PBS) and a half-wave plate (HWP). The beam impinged on the detector under test (MPPC), with the output connected to a data acquisition card (AD card). Calibration of the source was done in a Hanbury-Brown and Twiss (HBT) interferometer by erecting a flipping mirror (FM). The HBT consisted of a non-polarizing beamsplitter (NPBS) and two avalanche photodiodes (APD) connected to a coincidence circuit.

Independent calibration of the source was performed in a Hanbury-Brown and Twiss setup, with two APDs (Perkin-Elmer, SPCM-AQR-14), preceded by a non-polarizing beam splitter (NPBS). The APD outputs were addressed to a coincidence circuit (Acam messelectronic, TDC) with a 50 ns time window. From Eq. (4) one obtained $g_0^{(2)}=1.010\pm0.002$, which was then used in the fits of the experimental data (see Eq. (8)).

## 5. Results and discussion

First, the measurements described above were performed for three MPPCs with different pixel sizes. The resulting dependences $g^{(2)}(N_{Total,CT})$ for different detectors at -4.5ºC are shown in Fig.2. The crosstalk probabilities $p$, were obtained from fits of the experimental data which yielded typical values of coefficient of determination (COD) 0.983-0.99. For the sake of future comparison with the dark noise method of crosstalk calibration, we calculate the expression $p+2p^2$, which is shown in Table 1, as it represents the impact of crosstalk on the total number of counts in the non-linear model (see Eq. (7(b))).

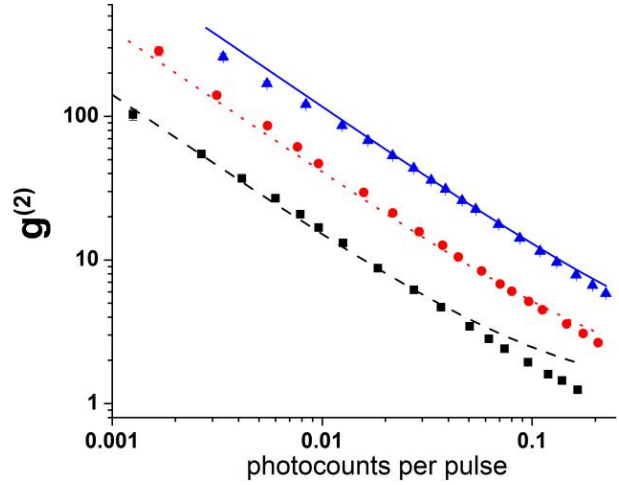

Fig.2 Dependences of $g^{(2)}$ on the mean number of photocounts per pulse $N_{Total,CT}$ measured at -4.5ºC by the MPPC with 25x25 μm² (black squares, dashed line), 50x50 μm² (red circles, dotted line), and 100x100 μm² pixel size (blue triangles, solid line). Lines are fits to experimental data, yielding COD 0.989, 0.983, and 0.997, for MPPC with 25x25 μm², 50x50 μm², and 100x100 μm² pixel size, respectively.

From the presented results it follows, those detectors with larger pixel size have larger crosstalk. This result is qualitatively confirmed by other groups using alternative calibration techniques, and explained by the dependence of the crosstalk probability on the gain of the detector [31]. The gain of each pixel is given by $G=C\,\Delta V$, where $C$ is the capacitance of the pixel and $\Delta V=V_{op}-V_{bd}$, where $V_{bd}$ is the breakdown voltage, and $V_{op}$ is the operational voltage. According to the datasheet, MPPC modules with larger size of the pixels have a larger gain, and consequently exhibit larger probability of crosstalk [17].

Additionally, the crosstalk was measured at 4 different temperatures of 25ºC, 12.5ºC, 5ºC, -4.5ºC for the MPPC with 50x50 μm² pixel size. The estimated crosstalk probabilities obtained from the fits of experimental dependences $g^{(2)}(N_{Total,CT})$ are summarized in Table 2, and shown in Fig.3 (red squares). The dependence of the crosstalk probability on the temperature is again attributed to the corresponding change of the gain [22]. Indeed, with the decrease of the temperature, $V_{bd}$ decreases, and leads to the increase of the gain. From the datasheet provided by the manufacturer, and our independent measurements, it was found that the MPPC module was designed to keep a constant $\Delta V$ by means of a compensation circuit, which regulates $V_{op}$ [17]. However, the obtained dependence of the crosstalk indicates, that for the tested MPPC modules the compensation circuit did not accurately follow the change of $V_{bd}$ with the temperature.

Finally, the results were compared with those obtained by the method based on the measurement of dark counts (see Eq. (2)). For each detector under test, dark counts were collected from the ensemble of $50*10^6$ triggers (see Table 1,2), which was intentionally chosen to be larger than the total amount of $36*10^6$ measurements used in calibration by $g^{(2)}$ measurements.

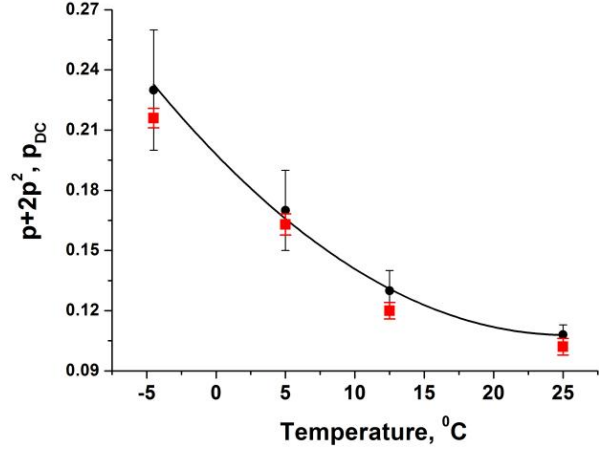

Fig. 3 Dependence of crosstalk probability $p+2p^2$ (red squares), found from $g^{(2)}$ measurements, and $p_{DC}$, measured from the dark noise (black circles) on temperature for MPPC with 50x50 μm² pixel size. Solid line is a fit with a quadratic function.

The results obtained by both methods are summarized in Tables 1 and 2, and shown in Fig.3. They show good agreement, with major deviations observed for the case of high crosstalk probability (large pixel size and low temperature). This is because the dark noise model does not account for nonlinear crosstalk terms, which contribution becomes non-negligible in this case. At the same time the method based on the measurement of $g^{(2)}$ gives the crosstalk probability with significantly less experimental uncertainty in the case of low dark counts (small pixel size and low temperature).

**Table 1. Probabilities of crosstalk measured for MPPCs with different pixel size at the temperature of -4.5ºC.**

| Pixel size, μm² | Dark noise, photocounts/pulse | $p+2p^2$ | $p_{DC}$ |
|---|---|---|---|
| 100x100 | 0.021 | 0.87±0.01 | 0.610±0.015 |
| 50x50 | 0.008 | 0.21±0.005 | 0.23±0.03 |
| 25x25 | 0.002 | 0.07±0.002 | 0.10±0.11 |

**Table 2. Probabilities of crosstalk measured for MPPC with 50x50 μm² pixel size at different temperatures.**

| Temperature, ºC | Dark noise, photocounts/pulse | $p+2p^2$ | $p_{DC}$ |
|---|---|---|---|
| 25 | 0.05 | 0.102±0.005 | 0.108±0.005 |
| 12.5 | 0.02 | 0.120±0.005 | 0.13±0.01 |
| 5 | 0.011 | 0.160±0.005 | 0.17±0.02 |
| -4.5 | 0.008 | 0.210±0.005 | 0.23±0.03 |

## 6. Conclusions

In conclusion a new accessible method for calibration of crosstalk has been presented and tested for MPPCs with different pixel sizes and at different temperatures. The method solely relies on the fundamental properties of coherent states, and the generalized crosstalk model. It does not require any *a-priori* knowledge of quantum efficiency of the detector under test. Obtained results were also compared with those measured by the conventional dark count method. Two methods were found to be in a good agreement qualitatively. At the same time, the new method exhibits much less uncertainty in determination of the crosstalk in the case of low dark counts. The beneficial aspects of the method are of a particular interest for practical applications, and future development of optimized low noise spatially resolved single photon detectors.

## Acknowledgement

We would like to acknowledge Maria Chekhova for stimulating discussions. This work was supported by A-STAR Investigatorship grant.